\UseRawInputEncoding 

\documentclass[aps,showpacs,onecolumn,notitlepage,floatfix,amsmath,amssymb,nofootinbib]{revtex4-1}

\usepackage{amssymb}
\usepackage{amsmath}
\usepackage{epsfig}
\usepackage{graphicx}
\usepackage{subfig}
\usepackage{color}
\usepackage{bigints}
\usepackage{latexsym}
\usepackage{bm,latexsym}
\usepackage{mathrsfs}
\usepackage{float}
\usepackage{pbox}
\usepackage[usenames,dvipsnames]{xcolor}
\definecolor{orange}{cmyk}{0,0.5,1,0}

\selectfont{}
\begin{document}

\title{ Novel traversable wormhole in General Relativity  and Einstein-Scalar-Gauss-Bonnet theory supported by nonlinear electrodynamics 
}

\author{Pedro Ca\~nate$^{\!^{1, 2}}$} 
\email[]{pcannate@gmail.com }
\author{F. H. Maldonado-Villamizar$^{\!^3}$}    
\email[]{fmaldonado@inaoep.mx }
\affiliation{$^{\!^1}$Departamento de F\'isica  Te\'orica, Instituto de F\'isica, Universidade do Estado do Rio de Janeiro, Rua S\~ao Francisco Xavier 524, Maracan\~a,
CEP 20550-013, Rio de Janeiro, Brazil. \\
$^{\!^2}$Programa de F\'isica, Facultad de Ciencias Exactas y Naturales, Universidad Surcolombiana, Avenida Pastrana Borrero - Carrera 1, A.A. 385, Neiva, Huila, Colombia.\\
%
%
$^{\!^3}$CONACYT-Instituto Nacional de Astrof\'isica, \'Optica y Electr\'onica,\\ Calle Luis Enrique Erro No. 1, Sta. Ma. Tonantzintla, Pue., 72840, Puebla, M\'exico}

\begin{abstract}
 Several traversable wormholes (T-WHs) of the Morris-Thorne type have been presented as exact solutions of Einstein-nonlinear electrodynamics gravity (GR-NLED), e.g. \cite{Arellano,Bronnikov2017,Bronnikov_Walia,Canate_Breton,Canate_Breton_Ortiz,Canate_Magos_Breton}.  
 However, none of these solutions is support by a nonlinear electrodynamics model satisfying plausible conditions. In this work, we present the first traversable wormhole solution of Einstein-nonlinear electrodynamics gravity coupled to a self-interacting phantom scalar field (GR-NLED-SF) 
 with a NLED model such that in the limit of weak field becomes the Maxwell electrodynamics, is presented. 
 Furthermore, we show that this novel T-WH spacetime is also an exact solution of the Einstein-scalar-Gauss-Bonnet (EsGB) theory with a nonlinear electrodynamics source, but now with a real scalar field having a positive kinetic term.
\end{abstract}



\maketitle

\section{Introduction}
The wormholes are  fascinating predictions arising from the geometrical description of gravity. They involve a topological spacetime configuration as a shortcut between distant points or regions in spacetime. The first wormhole interpretation originally came from the work of Einstein and Rosen in 1935, with their solution known as the Einstein-Rosen bridge \cite{einstein35}, which is, in essence, the maximally extended Schwarzschild black hole solution \cite{kruskal60}. However, the ``throat'' of this wormhole is dynamic and hence non-traversable, meaning that its radius expands to a maximum and quickly contracts to zero so fast that even a photon cannot pass through \cite{kruskal60}.
Further, in 1988, a solution to the wormhole traversability problem was established by Morris and Thorne  \cite{morris88}. They obtained one type of wormhole metric and the necessary conditions (absence of horizons and the flare-out condition) that can guarantee the traversability of a wormhole spacetime.
Moreover, they showed that in the context of general relativity (GR), the throat of these types of wormholes only could be kept open by some form of ``exotic'' matter \cite{morris88-2} having negative energy density and whose energy-momentum tensor violates the null-energy condition (NEC).
This kind of exotic fields have been intensely discussed,  
principally, in the context of electromagnetic and scalar fields 
\cite{Gibbons1996}.
Although an explanation about the fundamental origin of phantom fields 
is still in discussion, they are frequently used, for instance; 
in the construction of  novel 
cosmological models  known as Phantom Cosmologies \cite{Pha_Cos,Ph_SF1,Ph_SF}.  
Also, they have been used  to explain the current period of accelerated expansion
of the Universe.
Indeed, an accelerated expansion of the Universe can be accounted 
by assuming
the existence of an effective field generating repulsive gravity between its 
elements,
e.g., a repulsive field would correspond in Einstein gravity to a fluid with negative pressure, 
like  dark energy, which is an essential ingredient of the Standard Cosmological Model,  
 one of the most popular models in modern physics.  
A 
 source of repulsive gravity, in the context of exotic fields, would be represented
by a matter distribution with negative energy density, i.e.  a phantom field. In fact, comparison with observational data suggests it as 
a strong candidate for dark energy explanation  
\cite{Hannenstadt2006,Dunkle2009,Caldwell2003,Dutta2009,Alestas2020,Cedeno2021}.
Phantom fields are also considered 
in the astrophysics context. For instance,  Ref. \cite{Galactic} show 
the possibility that galactic dark matter exists
in a scenario where a phantom field is responsible for the dark energy distribution. \\
Therefore, is very important to understand the physical properties of phantom fields in the
framework of gravity theories.
In particular, a phantom scalar field could be used to generate the negative  kinetic energy density that allows traversable wormholes. 
For instance, one of the first and simplest examples of traversable wormholes is the static, spherically symmetric and asymptotically flat (SSS-AF) Ellis wormhole \cite{Ellis} whose energy-momentum tensor can be represented by a massless phantom scalar field. This solution has been extensively studied, and its properties like gravitational lensing \cite{EllisLensing}, quasi-normal modes \cite{QNMEllis}, shadows \cite{EllisShadows} and stability \cite{EllisStability} have been thoroughly investigated. Recently,  \cite{bronnikov13} shows that the source of the Ellis wormhole as a perfect fluid with negative energy density and a source-free radial electric or magnetic field is also possible.

The construction of traversable wormholes has been studied using non-linear electrodynamics (NLED) as source.
NLED  theories are derived from Lagrangians $\mathcal{L}=\mathcal{L}(\mathcal{F}, \mathcal{G})$ that depend arbitrarily on the two electromagnetic invariants, $\mathcal{F}= 2(\boldsymbol{\mathcal{E}}^2-\boldsymbol{\mathcal{B}}^2)$ and $\mathcal{G} = \boldsymbol{\mathcal{E}}\cdot\boldsymbol{\mathcal{B}}$,  where $\boldsymbol{\mathcal{E}}$ and $\boldsymbol{\mathcal{B}}$ are the electric and magnetic fields, respectively. Albeit this form is arbitrary, there exist two outstanding Lagrangians: 
 the Born-Infeld theory (BI)
\begin{equation}\label{BI}
\mathcal{L}_{_{\mathrm{BI}}}\!(\mathcal{F},\mathcal{G}) = 4b^{2} \left( -1 + \sqrt{ 1 + \frac{\mathcal{F}}{2b^{2}} + \frac{\mathcal{G}^{2}}{16b^{4}} }\right)
\end{equation}
where  $b$ is a constant which has the physical interpretation of a critical field
strength \cite{BI};
and the Euler-Heisenberg theory (EH), 
\begin{equation}\label{EH}
\mathcal{L}_{_{\mathrm{EH}}}\!(\mathcal{F},\mathcal{G}) = - \frac{1}{2}\mathcal{F} + \frac{\mu}{2}\mathcal{F}^{2} + \frac{7\mu}{8}\mathcal{G}^{2} 
\end{equation}
which corresponds to the weak field approximation of \cite{Heuler_35,Heisenberg_36},  and the coupling constant $\mu$ is written as $\mu=2\alpha^{2}/(45m^{4}_{e})$, where $m_{e}$ is the mass of the electron and $\alpha$ is the fine structure constant. 
Considering  NLED Lagrangian  of the form $\mathcal{L}(\mathcal{F})$  coupled to gravity, interesting solutions arise, like regular black holes or traversable wormholes, among others  \cite{Ayon_Garcia,Arellano,Bronnikov2017,EH_BH,Novello}. 

By using NLED as a source in the Einstein field equations, two kind of T-WH have been investigated: 
dynamic \cite{Arellano,Bronnikov2017} and static \cite{Bronnikov_Walia}. While, dynamic T-WHs are possible in the GR context by using NLED as the only source, the SSS T-WHs static wormholes are not possible in the NLED context \cite{Arellano2006}. To date, the  SSS T-WHs studied requires an additional scalar field \cite{Bronnikov_Walia}.
Howerver, the common NLED models used for the construction of dynamical and static traversable wormholes do not satisfy plausible physical conditions, like how to reduce them to Maxwell electrodynamics  in the weak field limit \cite{Bronnikov2017,Bronnikov_Walia,Canate_Breton,Canate_Breton_Ortiz,Canate_Magos_Breton}.  

In this paper we will construct a SSS-AF T-WH, to do this we consider a Euler-Heisenberg-like electrodynamics model as the NLED source, with the Lagrangian density defined by:
\begin{equation}\label{EHT}
\mathcal{L}(\mathcal{F}) = - \frac{1}{2}\mathcal{F} + \mu_{_{0}} \mathcal{F}^{^{2}}  + \mu_{_{1}} |\mathcal{F}|^{^{\frac{3}{2}}},
\end{equation}
where $\mu_{_{0}}$ and $\mu_{_{1}}$ are real parameters of the model. 
Moreover, we use a self-interacting  scalar field, described by the  potential,
\begin{equation}
\mathscr{U}(\phi) = \mathscr{U}_{_{0}} + \frac{\beta_{_{0}}}{2}\!\!\left(\frac{\phi}{2}\right)^{\!\!\!^{4}} + \frac{2\beta_{_{1}}}{3}\!\!\left(\frac{\phi}{2}\right)^{\!\!\!^{6}} + \frac{\beta_{_{2}}}{4}\!\!\left(\frac{\phi}{2}\right)^{\!\!\!^{8}},      
\end{equation}
being $\mathscr{U}_{_{0}} = \mathscr{U}(0)$, whereas $\beta_{_{0}}$, $\beta_{_{1}}$ and $\beta_{_{2}}$ are real parameters.  
This power-law potential has interesting applications in cosmology \cite{Ramirez,Senoguz,Amit,Takyi}.

This paper is structured as follows: In Section \ref{FieldEqs}  we derive the field equations for the GR-NLED-SF, in Section \ref{Canonical_TWH} we present the canonical metric of the traversable wormhole spacetime and discuss the Ellis wormhole solution, in Section \ref{New_TWH} we construct our novel traversable wormhole solution  within the framework of GR-SF-NLED. Whereas, in Section \ref{ESTGBgravity}, we show that our T-WH spacetime is also a exact solution of EsGB-NLED, but now with a real
scalar field having a positive kinetic term. 
In Section \ref{Nullgeods} the null trajectories and the capture cross-section of massless (photon) by this T-WH spacetime, are analyzed.  
In the last section final conclusions are presented. Through this paper we will use the system of units where $G = k_{_{B}} = c =\hbar = 1$, and the metric signature  $(-+++)$.

\section{Einstein-nonlinear electrodynamics gravity coupled to a self-interacting scalar field 
}\label{FieldEqs}

The GR-NLED-SF theory is defined by the following action,
\begin{equation}\label{action_STNLED}
S[g_{\alpha\beta},\phi,A_{\nu}] = \int d^{4}x \sqrt{-g} \left\{ \frac{1}{16\pi}\left(R - \frac{1}{2}\partial_{\mu}\phi\partial^{\mu}\phi  - 2 \mathscr{U}(\phi) \right) + \frac{1}{4\pi}\mathcal{L}(\mathcal{F}) \right\},
\end{equation}
where $R$ is the scalar curvature, $\phi$ is a scalar field coupled to gravity, and  $\mathscr{U}=\mathscr{U}(\phi)$ is the scalar potential; whereas $\mathcal{L}=\mathcal{L}(\mathcal{F})$ is a function of the electromagnetic invariants $\mathcal{F}\equiv \frac{1}{4}F_{\alpha\beta}F^{\alpha\beta}$, being $F_{\alpha\beta}=2\partial_{[\alpha}A_{\beta]}$  
the components of the electromagnetic field tensor  $\boldsymbol{F}=\frac{1}{2}F_{\alpha\beta} \boldsymbol{dx^{\alpha}} \wedge \boldsymbol{dx^{\beta}}$ and $A_{\alpha}$ are the components of the electromagnetic potential. 

Using the notation $\mathcal{L}_{_{\mathcal{F}}}\equiv \frac{d\mathcal{L}}{d\mathcal{F}}$ and $\dot{\mathscr{U}} = \frac{d\mathscr{U}}{d\phi}$, the GR-NLED-SF field equations arising from  (\ref{action_STNLED}) takes the form, 
\begin{equation}\label{ES_NLED_Eqs}
G_{\alpha}{}^{\beta} = 8\pi (E_{\alpha}{}^{\beta})\!_{_{_{S \! F}}} \!+ \! 8\pi (E_{\alpha}{}^{\beta})\!_{_{_{N \! L \! E \! D}}}, \quad\quad\quad 
\nabla_{\mu}(\mathcal{L}_{_{\mathcal{F}}}F^{\mu\nu}) = 0 = d\boldsymbol{F},  
\quad\quad\quad \nabla^{2}\phi = 2\hspace{0.03cm}\dot{\mathscr{U}}, %
\end{equation}
where, $G_{\alpha}{}^{\beta} = R_{\alpha}{}^{\beta} - \frac{R}{2} \delta_{\alpha}{}^{\beta}$  are the components of the Einstein tensor,  $(E_{\alpha}{}^{\beta})\!_{_{_{S \! F}}}$ are the components of the energy-momentum tensor of self-interacting scalar field, 
\begin{equation}\label{E_SF}
8 \pi(E_{\alpha}{}^{\beta})\!_{_{_{S \! F}}} = -\frac{1}{4}(\partial_{\mu}\phi \hspace{0.04cm} \partial^{\mu}\phi)\delta_{\alpha}{}^{\beta} + \frac{1}{2}\partial_{\alpha}\phi \hspace{0.04cm} \partial^{\beta}\phi - \mathscr{U} \hspace{0.03cm} \delta_{\alpha}{}^{\beta},     
\end{equation}
and $(E_{\alpha}{}^{\beta})\!_{_{_{N \! L \! E \! D}}}$  are the components of the NLED energy-momentum tensor,
\begin{equation}\label{NLED_EM}
8\pi(E_{\alpha}{}^{\beta})\!_{_{_{N \! L \! E \! D}}} = 2\mathcal{L}_{_{\mathcal{F}}}F_{\alpha\mu}F^{\beta\mu} - 2\mathcal{L}\hskip.06cm\delta_{\alpha}{}^{\beta}. %
\end{equation}

Our aim is to find a static, spherically symmetric, and asymptotically flat, charged wormhole solution  for the set  Eqs. (\ref{ES_NLED_Eqs}) with a non-trivial scalar field. To do this, we assume  the scalar field is static and spherically symmetric, $\phi = \phi(r)$, and also  the metric has the static and spherically symmetric form,
\begin{equation}\label{SSSmet}
ds^{2} =  - e^{ A(r) }dt^{2} + e^{ B(r) }dr^{2}  + r^{2}(d\theta^{2}  + \sin^{2}\theta d\varphi^{2}),  
\end{equation}
with $A = A(r)$ and $B = B(r)$ unknown functions to be determined.

In terms of $A$ and $B$, the non-vanishing  components of the Einstein tensor  are,
\begin{equation}\label{GabSSS}
G_{t}{}^{t}\!=\!\frac{ e^{^{\!\!-B}} }{ r^{2} }\!\!\left( -rB' - e^{^{\!B}} + 1 \right)\!, \hspace{0.25cm}  G_{r}{}^{r} \!=\! \frac{ e^{^{\!\! -B}} }{ r^{2} }\!\!\left( rA' - e^{^{\!B}} + 1 \right)\!, \hspace{0.25cm} G_{\theta}{}^{\theta}\!=\! G_{\varphi}{}^{\varphi}\!=\!\frac{ e^{^{\!\!-B}} }{ 4r }\!\!\left( rA'^{2} - rA'B' + 2rA'' + 2A' - 2B' \right)\!,
\end{equation}
where $'$ denotes the  derivative respect to the radial coordinate $r$, i.e. $A' = \frac{dA}{dr}$.
Whereas, for the non-trivial components of the energy-momentum tensor of self-interacting scalar field we have,
\begin{equation}\label{EttyErr}
8 (E_{t}{}^{t})_{\!_{SF}} = 8\pi (E_{\theta}{}^{\theta})_{\!_{SF}} = 8\pi (E_{\varphi}{}^{\varphi})_{\!_{SF}} = -\frac{1}{ 4 } e^{ -B} \phi'^{2}  - \mathscr{U}, \quad\quad\quad\quad 8\pi (E_{r}{}^{r})_{\!_{SF}} = \frac{1}{ 4 } e^{ -B} \phi'^{2} - \mathscr{U}. 
\end{equation}

Regarding the electromagnetic field tensor, since the spacetime is SSS, we can restrict ourselves to purely magnetic field, i.e., $\mathcal{E} = 0$ and $\mathcal{B} \neq 0$.

 With this restriction, the electromagnetic field tensor has the form, $F_{\alpha\beta} = \mathcal{B}\left( \delta^{\theta}_{\alpha}\delta^{\varphi}_{\beta} - \delta^{\varphi}_{\alpha}\delta^{\theta}_{\beta} \right)$.  
In this way, for a static and spherically symmetric spacetime with line element (\ref{SSSmet}), the general solution of the equations $\nabla_{\mu}(\mathcal{L}_{_{\mathcal{F}}}F^{\mu\nu})=0$ is,
\begin{equation}\label{fabSOL}
F_{\theta\varphi} =  r^{4} \mathcal{Q}(r) \sin\theta.
\end{equation}
This means  $\boldsymbol{F} = r^{4} \mathcal{Q}(r)\sin\theta \hspace{0.1cm} \boldsymbol{d\theta} \wedge \boldsymbol{d\varphi}$, and therefore $d\boldsymbol{F} = 0 = (r^{4} \mathcal{Q}(r))' \sin\theta \hspace{0.1cm} \boldsymbol{dr} \wedge \boldsymbol{d\theta} \wedge \boldsymbol{d\varphi}$. This implies \begin{math}\mathcal{Q}(r)~=~\sqrt{2}\hskip.06cm q/r^{4}\end{math}, where
$\sqrt{2}\hskip.06cm q$ is an integration constant,  which  plays the role of the magnetic charge.
Hence, the components of the electromagnetic field tensor $F_{\alpha \beta}$ and the invariant $\mathcal{F}$ are
given by;
\begin{equation}\label{magnetica}
F_{\alpha\beta} = \sqrt{2} \hspace{0.05cm} q  \sin \theta \hspace{0.05cm} \left( \delta^{\theta}_{\alpha}\delta^{\varphi}_{\beta} - \delta^{\varphi}_{\alpha}\delta^{\theta}_{\beta} \right),  \quad\quad\quad\quad \mathcal{F} =  \frac{q^{2}}{r^{4}}.
\end{equation}

Finally, the  energy-momentum tensor components 
for NLED,  assuming the SSS spacetime with metric (\ref{SSSmet}), the purely magnetic field  (\ref{magnetica}), and a generic Lagrangian density $\mathcal{L}(F)$, are written as
\begin{eqnarray}\label{E_nled}
8\pi (E_{t}{}^{t})\!_{_{_{N \! L \! E \! D}}} = 8\pi (E_{r}{}^{r})\!_{_{_{N \! L \! E \! D}}} =  -2\mathcal{L}, 
\quad\quad\quad 8\pi (E_{\theta}{}^{\theta})\!_{_{_{N \! L \! E \! D}}} = 8\pi (E_{\varphi}{}^{\varphi})\!_{_{_{N \! L \! E \! D}}} =  2(2\mathcal{F}\mathcal{L}_{\mathcal{F}} - \mathcal{L}). 
\end{eqnarray}
%
Inserting the above  components in the field equations
$C_{\alpha}{}^{\beta} = G_{\alpha}{}^{\beta} - 8\pi [ (E_{\alpha}{}^{\beta})\!_{_{_{S \! F }}} + (E_{\alpha}{}^{\beta})\!_{_{_{N \! L \! E \! D}}} ] = 0$, we obtain the GR-NLED-SF field equations for the metric ansatz (\ref{SSSmet}) and the magnetic field (\ref{magnetica}):
\begin{eqnarray}
&&\!C_{t}{}^{t}\!=\!0\hspace{1.4cm}\!\Rightarrow\!\hspace{0.4cm}
\frac{ e^{^{\!\!-B}} }{ r^{2} }\!\!\left( -rB' - e^{^{\!B}} + 1 \right) + \frac{1}{ 4 } e^{ -B} \phi'^{2}  + \mathscr{U} + 2\mathcal{L} =0, 
\label{AB_Eqt}\\
&&\nonumber\\
&&\!C_{r}{}^{r}\!=\!0\hspace{1.4cm}\!\Rightarrow\!\hspace{0.4cm} \frac{ e^{^{\!\! -B}} }{ r^{2} }\!\!\left( rA' - e^{^{\!B}} + 1 \right) - \frac{1}{ 4 } e^{ -B} \phi'^{2} + \mathscr{U} + 2\mathcal{L} = 0,
\label{AB_Eqr}\\
&&\nonumber\\
&&\!C_{\theta}{}^{\theta}\!=\!C_{\varphi}{}^{\varphi}\!=\!0\hspace{0.4cm}\!\Rightarrow\!\hspace{0.4cm} 
\frac{ e^{^{\!\!-B}} }{ 4r }\!\!\left( rA'^{2} - rA'B' + 2rA'' + 2A' - 2B' \right) + \frac{1}{ 4 } e^{ -B} \phi'^{2}  + \mathscr{U} - 2(2\mathcal{F}\mathcal{L}_{\mathcal{F}} - \mathcal{L})  =0.
\label{AB_Eqte}
\end{eqnarray}
whereas the scalar field must satisfies,
\begin{equation}\label{AB_phi2}
2r\phi'' + (4 + rA' - rB')\phi' - 4r e^{B}\dot{\mathscr{U}} = 0.  
\end{equation}
This ends the general treatment of the static, spherically symmetric and pure magnetic solutions. In what follows we will discuss the general properties of the wormhole spacetime solution we have obtained.

\section{The canonical metric of a wormhole spacetime and traversability}\label{Canonical_TWH} 
The canonical metric of a (3+1)-dimensional SSS-WH  solution \cite{morris88,morris88-2} is given by,
\begin{equation}\label{TWH_MT}
\boldsymbol{ds^{2}}_{_{W\!H}} = - e^{2\Phi(r)}\boldsymbol{dt^{2}} + \frac{\boldsymbol{dr^{2}}}{ 1 - \frac{b(r)}{r}} + r^{2}\boldsymbol{d\Omega^{2}},
\end{equation}  
where $\Phi(r)$ and $b(r)$ are smooth functions, known as redshift and shape functions respectively. 
The domain for radial coordinate $r$ has a  minimum  at $r=r_{_{0}}$, where the WH throat is defined by $b(r_{_{0}}) = r_{_{0}}$, and is unbounded for $r>r_{_{0}}$. This coordinate has a special geometric interpretation, as  $4\pi r^2$  is the area of a sphere centered on the WH throat.
On the other hand, for the WH to be traversable, one must demand:
\begin{eqnarray} 
&&\textup{{\bf Wormhole domain:}}\hspace{2.8cm}   
 1 - \frac{b(r)}{r} > 0  \quad\quad\quad\quad \forall\quad\!\!\!\!\!r>r_{_{0}}  \label{TWC1}\\
&& \textup{{\bf Absence of horizons:}}\hspace{2.5cm}  e^{2\Phi(r)}\in\mathbb{R}^{+}\!-\!\{0\}  \quad\quad  \forall\quad\!\!\!\!\! r\geq r_{_{0}} , \quad\quad \textup{ and } \quad\quad \Phi(r\rightarrow\infty)=0\label{TWC2}\\
&&\textup{{\bf Flare-out condition:}}\hspace{2.6cm} b'(r)\Big|_{r=r_{_{0}}}<1\label{TWC3}
\end{eqnarray}
with $'$ denoting derivative with respect to $r$,
are satisfied (see \cite{morris88,morris88-2} for details). 

{\bf Traversability and violation of Null Energy Condition.} \\\noindent
Let us consider the null vector $\boldsymbol{n}~=~( e^{-\Phi(r)}, \sqrt{ 1  - b(r)/r }, 0, 0 )$~, identify  $e^{A(r)} ~=~e^{2\Phi(r)}$  and $e^{B(r)} = \frac{1}{1 - \frac{b(r)}{r} }$ in the spacetime (\ref{TWH_MT}). Using  (\ref{GabSSS})
and assuming  the flaring out condition is fulfilled,  after contracting the Einstein tensor with $\boldsymbol{n}$ and evaluating at $r=r_{_{0}}$, yields:
\begin{eqnarray}\label{Gnnr0}
 G_{\alpha\beta}n^{\alpha}n^{\beta}  \Big|_{r=r_{_{0}}} = \left(G_{r}{}^{r} - G_{t}{}^{t}\right)\Big|_{r=r_{_{0}}} 
 =\frac{1}{r^{2}_{_{0}}} \left[ b'(r_{_{0}})- 1  \right]< 0. 
\end{eqnarray}
Thus, in GR, $G_{\alpha\beta} = \kappa T_{\alpha\beta}$, from the balance between the matter and the curvature quantities, the fulfillment of the flaring out condition implies that the NEC (which states that $T_{\alpha\beta}n^{\alpha}n^{\beta}\geq 0$ for any null vector $n^{\alpha}$) is violated,  therefore,   the presence of exotic matter is unavoidable for having a T-WH in GR.

\subsection{  Ultra-static spherically symmetric and asymtotically flat solution in GR-NLED-massless scalar field theory: The Ellis wormhole } 
A spacetime is called ultra-static if it admits an atlas of charts in which the metric tensor takes the
form,
\begin{equation}\label{UltraS}
\boldsymbol{ds^{2}} = - \boldsymbol{dt^{2}} + g_{ab}\boldsymbol{dx^{a}}\boldsymbol{dx^{b}} ,
\end{equation}  
where the  metric coefficients $g_{ab}$ are independent of the time coordinate $t$, and in where the Latin indices running over the spatial coordinates only.
These spacetimes have interesting properties \cite{Fulling,Fulling81,DonPage}.
Setting $A(r)=0$ in (\ref{SSSmet}) one arrive to the canonical metric for the  ultra-static spherical symmetric spacetime.  \\
Now, if we consider GR-NLED-SF with $\mathscr{U}(\phi)=0$ and with $\mathcal{E}\neq 0$ and $\mathcal{B}\neq 0$ see Appendix \ref{EB_FieldEqs},    
the   equation of motion for the scalar field (\ref{AB_phi2}) yields, 
\begin{equation}\label{prof1}
2r\phi'' + (4 - rB')\phi' = 0 \quad\quad\Rightarrow\quad\quad \phi'^{2} = \frac{ \gamma e^{^{B}} }{r^{4}}  \quad\quad\textup{being}\quad\quad \gamma = \textup{constant} \in\mathbb{R}.  
\end{equation}
 Subtracting (\ref{AB_Eqr_EM})  from  (\ref{AB_Eqt_EM}), yields, 
\begin{equation}\label{prof2}
\phi'^{2} = \frac{ 2B'}{ r }.
\end{equation}
Now, equating  (\ref{prof1}) with (\ref{prof2}), gives,
\begin{equation}\label{prof3}
\frac{ \gamma e^{^{B}} }{2r^{3}B'} = 1.
\end{equation}
 Identifying (\ref{SSSmet}) with (\ref{TWH_MT}),  yields $A = 2\Phi(r)=0$, $B = - \ln\left(1 - \frac{b(r)}{r}\right)$  and  substituting them in (\ref{prof3}) yields,
\begin{equation}\label{prof5}
\frac{ \gamma  }{2r^{2}\left[ b' - \frac{b}{r} \right] } = 1.      
\end{equation}
 Evaluating in the WH throat yields,  
\begin{equation}\label{prof6}
\frac{ \gamma  }{2r^{2}_{_{0}}\left[ b'(r_{_{0}}) - 1 \right] } = 1 \quad\quad\textup{{\bf and}}\quad\quad b'(r_{_{0}})<1 \quad\textup{({\bf flare-out condition })} \quad\Rightarrow\quad \gamma = - 4 q^{2} < 0.  
\end{equation}
Replacing $\gamma$  in (\ref{prof5}) and solving for $b$, gives
\begin{equation}\label{prof7}
b =   \tilde{q}\!~r + \frac{q^{2}}{r} \quad\textup{being}\quad \tilde{q} = \textup{constant} \in\mathbb{R}.
\end{equation}
Hence, in this case the metric (\ref{TWH_MT}) has the form, 
\begin{equation}\label{prof8}
ds^{2} =  - dt^{2} + \frac{dr^{2}}{ 1 - \tilde{q} - \frac{q^{2}}{r^{2}} }   + r^{2}(d\theta^{2}  + \sin^{2}\theta d\varphi^{2}).
\end{equation}
In order to this metric be asymptotically flat is necessary $\tilde{q}=0$, and then we arrive to the ultra-static SS-AF metric given by,
\begin{equation}\label{EllisWH}
ds^{2} =  - dt^{2} + \frac{dr^{2}}{ 1 - \frac{q^{2}}{r^{2}} }  + r^{2}(d\theta^{2}  + \sin^{2}\theta d\varphi^{2}).
\end{equation}
Then, for this spacetime metric the solution of (\ref{AB_phi2}) is 
\begin{equation}\label{prof9}
\phi(r) = \phi_{_{\ast}} + 2i\tan^{^{\!\!-1}}\!\!\left( \sqrt{\frac{r^{2} - q^{2}}{q^{2}}}\right),
\end{equation}
where $\phi_{_{\ast}}$ is a integration constant which can be fixed to zero without loss generality.
By substituting (\ref{prof8}) and (\ref{prof9}) in (\ref{AB_Eqt_EM}), (\ref{AB_Eqr_EM}) and (\ref{AB_Eqte_EM})  with $\mathscr{U}(\phi)=0$, yields $F_{tr}F^{tr}\mathcal{L}_{\mathcal{F}} - \mathcal{L}  = 0$ and  $F_{\theta\phi}F^{\theta\phi}\mathcal{L}_{\mathcal{F}} - \mathcal{L}= 0$, these equations imply.  
\begin{equation}\label{FtrFthph}
F_{tr}F^{tr} = F_{\theta\phi}F^{\theta\phi},    
\end{equation}
On the other hand we can calculate the quantities $F_{tr}F^{tr}$ and $F_{\theta\phi}F^{\theta\phi}$, obtaining in the spacetime region $r>|q|$,  
\begin{equation}
F_{tr}F^{tr} = g^{tt}g^{rr}(F_{tr})^{2} = - \left( 1 - \frac{q^{2}}{r^{2}} \right)(F_{tr})^{2} < 0 \quad\quad\textup{and}\quad\quad F_{\theta\phi}F^{\theta\phi} = g^{\theta\theta}g^{\phi\phi}(F_{\theta\phi})^{2} = \frac{(F_{\theta\phi})^{2}}{r^{4}\sin^{2}\theta}  > 0.
\end{equation}
Then, the equation (\ref{FtrFthph}) is valid only  if $F_{tr} = F_{\theta\phi} = 0$, this means the electromagnetic field  and the associated Lagrangian energy density  $\mathcal{L}$ must be both  zero.
\\
The metric (\ref{EllisWH}), originally introduced in \cite{Ellis}, admits a T-WH interpretation since satisfies the properties (\ref{TWC1})-(\ref{TWC3}), and is known as the Ellis wormhole metric.

Indeed, defining a new scalar field by $\psi = i\phi$ (phantom field), and using $\mathscr{U}(\phi)=\mathcal{L}(\mathcal{F})=0$ in the action   
(\ref{action_STNLED}), the wormhole metric (\ref{EllisWH}) becomes a solution to the theory
with gravitational action,
\begin{equation}\label{actionL2}
S[g_{\alpha\beta},\psi] = \int d^{4}x \sqrt{-g} \left\{ \frac{1}{16\pi}\left(R + \frac{1}{2}\partial_{\mu}\psi\partial^{\mu}\psi \right)  \right\},
\end{equation}
with $\psi$ given by,
\begin{equation}
\psi =  2\tan^{-1}\left( \sqrt{ \frac{r^{2}  - q^{2}}{q^{2}} } \right)  \in \mathbb{R}.
\end{equation}
This is the action that was used by Ellis in Ref. \cite{Ellis} to get the ultra-static wormhole solution (\ref{EllisWH}).

\section{Pure-magnetic T-WH  supported by NLED in Einstein-scalar gravity: A non-ultra-static modification of Ellis wormhole}\label{New_TWH}

In this section we will study the  NLED-SF theory defined  by NLED model and a scalar potential, given respectively by,
\begin{eqnarray}
 \mathcal{L}(\mathcal{F}) &=& - \frac{1}{2}\mathcal{F} + \mu_{_{0}} \mathcal{F}^{^{2}}  
 + \mu_{_{1}} |\mathcal{F}|^{^{\frac{3}{2}}},  %
 \label{NLmax1}
\\
\mathscr{U}(\phi) &=& \mathscr{U}_{_{0}} + \frac{\beta_{_{0}}}{2}\!\!\left(\frac{\phi}{2}\right)^{\!\!\!\!^{4}} + \frac{2\beta_{_{1}}}{3}\!\!\left(\frac{\phi}{2}\right)^{\!\!\!\!^{6}} + \frac{\beta_{_{2}}}{4}\!\!\left(\frac{\phi}{2}\right)^{\!\!\!\!^{8}}, %
\label{non_ELLis_model_U}  
\end{eqnarray}
admits the following metric,
\begin{equation}\label{non_ELLis_WH}
\boldsymbol{ds^{2}} = - e^{\!^{ -\frac{q^{2}}{r^{2}} } }\boldsymbol{dt^{2}} + \frac{\boldsymbol{dr^{2}}}{\big(1 - \frac{q^{2}}{r^{2}} \big)} + r^{2}\boldsymbol{d\Omega^{2}} 
\end{equation}
for $\mu_{_{0}}=-q^{2}/8$, $\mu_{_{1}}=2|q|/3$, $\mathscr{U}_{_{0}}=-1/(12q^{2})$, $\beta_{_{0}} = \beta_{_{1}} = \beta_{_{2}} = 1 /q^{2}$, together with the scalar field,
\begin{equation}\label{SF_non_Ellis}
\phi(r) = 2i \sqrt{ 1 - \frac{q^{2}}{r^{2}} }     
\end{equation} 
as a pure magnetic exact solution of the GR-NLED-SF field equations (\ref{AB_Eqt})-(\ref{AB_phi2}). 
The metric (\ref{non_ELLis_WH}) admits a T-WH interpretation since satisfies the properties (\ref{TWC1})-(\ref{TWC3}), because this metric has the form (\ref{EllisWH}) but with non vanishing redshift function given by $\Phi(r) = -\frac{q^{2}}{r^{2}}$, can be interpreted as a non-trivial redshift modification of the Ellis WH metric, with WH throat at $r=r_{_{0}}=|q|$.
In other words, the line element  (\ref{non_ELLis_WH}) is not of type (\ref{UltraS}), therefore it describes a non-ultra-static spherically symmetric asymptotically flat traversable wormhole.

The absence of curvature singularities in the WH domain, $r\geq|q|$, 
can be deduced from the analytical expressions of its curvature invariants,
\begin{eqnarray}
&&R = \frac{ 2q^{4}\left( q^{2}-3r^{2}\right)}{r^{8}},  \quad\quad\quad\quad
R_{\alpha\beta}R^{\alpha\beta} =  \frac{ 2q^{4}\left(q^{8}-8q^{6}r^{2} + 20q^{4}r^{4} - 10q^{2}r^{6} +2r^{8}\right)}{r^{16}}, \label{R_R2R2} %
\\
&&R_{\alpha\beta\mu\nu}R^{\alpha\beta\mu\nu} = \frac{ 4q^{4}\left[ q^{8}-10q^{6}r^{2} + 33q^{4}r^{4} - 34q^{2}r^{6} + 14r^{8}   \right]}{r^{16}}, \label{R4R4}
\end{eqnarray} 
which are all regular in the whole WH domain $r\geq |q|$ .    

Here, it is important to emphasise that the Lagrangian density (\ref{NLmax1})   
reduces to Maxwell theory in the limit of weak field, i.e. $\mathcal{L} \rightarrow \kappa \mathcal{F} $ and $\mathcal{L}_{\mathcal{F}} \rightarrow \kappa $ (being $\kappa$ a constant) as $\mathcal{F}\rightarrow0$.


Besides the Maxwell limit condition, another important physical requirement is desired for the electromagnetic  Lagrangian (\ref{NLmax1})  to fulfill, it is the  
WEC. We can guarantee the validity of WEC in a limited region of the spacetime determined by,
\begin{equation}
\frac{92 - 32\sqrt{7}}{9} \leq q^{2}\mathcal{F} \leq 1  \quad\quad \Rightarrow \quad\quad  \frac{r}{|q|} \in    \left[1 , \left(\frac{9}{92 - 32\sqrt{7}}\right)^{\!\!^{\frac{1}{4}}}  \right]
\end{equation}
which includes the wormhole throat $r=r_{_{0}}=|q|$. See Appendix \ref{WEC_NEC_Model} for details.

{\bf Phantom scalar field:}
Defining a new scalar field by $\psi = i\phi$ (phantom field), the theory for the which the metric (\ref{non_ELLis_WH}) is a pure magnetic exact solution, arises from the action,   
\begin{equation}
S[g_{\alpha\beta},\psi,A_{\nu}] = \int d^{4}x \sqrt{-g} \left\{ \frac{1}{16\pi}\left(R + \frac{1}{2}\partial_{\mu}\psi\partial^{\mu}\psi  - 2 \mathscr{U}(\psi) \right) + \frac{1}{4\pi}\mathcal{L}(\mathcal{F}) \right\},
\end{equation}
with nonlinear electromagnetic field described by (\ref{NLmax1}) and  
and scalar field potential,  
\begin{equation}
\mathscr{U}(\psi) = -\frac{\beta_{_{0}}}{48} 
\left(3\psi^{2} + 4\right)\left( 1 - \frac{\psi^{2}}{4} \right)^{\!\!\!^{3}}, 
\end{equation}
with $\psi$ given by, 
\begin{equation}
\psi(r) = 2\sqrt{ 1 - \frac{q^{2}}{r^{2}} }  \in \mathbb{R}  
\end{equation}
    

\section{Einstein-scalar-Gauss-Bonnet theory} \label{ESTGBgravity} 
Einstein-scalar-Gauss-Bonnet (EsGB) theories belong to a class of alternative theories of gravity, in which the Einstein-Hilbert action with scalar field, $\phi$, that is (\ref{action_STNLED}), is modified by including a  
quadratic curvature correction given by the product of a function of
the scalar field, $\boldsymbol{f}(\phi)$, and the  Gauss-Bonnet term $R_{_{GB}}^{2}=R_{\alpha\beta\mu\nu}R^{\alpha\beta\mu\nu} - 4R_{\alpha\beta}R^{\alpha\beta} + R^{2}$.
Thus, the dynamical equations of EsGB theory minimally coupled to matter fields  
are derived 
from the following action,
\begin{equation}\label{actionL}
S[g_{ab},\phi,\psi_{a}] = \int d^{4}x \sqrt{-g} \left\{ \frac{1}{16\pi}\left(R - \frac{1}{2}\partial_{\mu}\phi\partial^{\mu}\phi - 2 \mathscr{U}(\phi) \right) + \frac{1}{16\pi} \boldsymbol{f}(\phi) R_{_{GB}}^{2} + \frac{1}{4\pi}
\mathcal{L}_{\rm matter}(g_{ab},\psi_{a})  
\right\},
\end{equation}
where $\boldsymbol{f}(\phi)R_{_{GB}}^{2}$ stands for the scalar field non-minimally coupled to the Gauss-Bonnet invariant, or Scalar field-Gauss-Bonnet (SFGB) term, 
$\mathcal{L}_{\rm matter}$ is the matter fields Lagrangian and $\psi_{a}$ describes any matter field included in the EsGB theory.
Although the Gauss-Bonnet term includes quadratic curvature components, the resulting field equations arising from it 
are of second order, and therefore avoid the Ostrogradski instability and ghosts. 
Various aspects of the EsGB theory
have been studied in detail lately. To name just a few applications in cosmology, 
it was shown in 
\cite{cosmic_acceleration} that 
the theory can describe 
the
present stage of cosmic acceleration, and
can provide an exit from a
scaling matter-dominated epoch to a late-time accelerated expansion \cite{Tsujikawa2006}. The possible reconstruction of the coupling and potential functions for a given scale factor was considered in \cite{recon}, and
the consequences of the EsGB in an inflationary setting have been considered in \cite{Odintsov2020}.
In our case the coupling between the scalar field and the Gauss-Bonnet term  
allows the violation of the null-energy condition required for the traversable Ellis wormhole, and so in a sense the effective negative energy density comes from the geometry itself instead of the matter source as in GR. Recently, in  \cite{antoniou19} various 
novel wormhole solutions in EsGB theory have been obtained, numerically, for several coupling functions. Other numerical traversable wormhole solutions have been obtained earlier \cite{kanti11} in Einstein-dilaton-Gauss-Bonnet theory \cite{kanti96}, which involve an exponential coupling between the scalar field representing the dilaton and the Gauss-Bonnet term.
Exact traversable wormhole solutions have been discussed within EsGB, however, at present only 
ultra-static are available \cite{canate19,canate19b}.
From now on, we will consider the EsGB theory in the presence of  
NELD described by  
$\mathcal{L}(\mathcal{F})$.
The EsGB-NLED field equations arising from
the action (\ref{actionL}) with $\mathcal{L}_{\rm matter}=\mathcal{L}(\mathcal{F})$,
are given by, 
\begin{equation}\label{modifEqs}
G_{a}{}^{b} = 8\pi (E_{a}{}^{b})\!_{_{_{S\!F\!G\!B}}} + 8\pi (E_{a}{}^{b})\!_{_{_{N \! L \! E \! D}}}, \quad\quad\quad 
\nabla_{\mu}(\mathcal{L}_{_{\mathcal{F}}}F^{\mu\nu}) = 0 = d\boldsymbol{F},  
\quad\quad\quad \nabla^{2}\phi + \dot{\boldsymbol{f}}(\phi) R_{_{GB}}^{2} + 2 \dot{\mathscr{U}}(\phi) = 0, 
\end{equation}
where $(E_{a}{}^{b})\!_{_{_{S\!F\!G\!B}}}$,  
denotes the components of tensor,  
\begin{eqnarray}
&& 8\pi (E_{\alpha}{}^{\beta})\!_{_{_{S\!F\!G\!B}}} = -\frac{1}{4}(\partial_{\mu}\phi \partial^{\mu}\phi)\delta_{\alpha}{}^{\beta} + \frac{1}{2}\partial_{\alpha}\phi \partial^{\beta}\phi - \mathscr{U}(\phi)\delta_{\alpha}{}^{\beta} - \frac{1}{2}( g_{\alpha\rho} \delta_{\lambda}{}^{\beta} + g_{\alpha\lambda} \delta_{\rho}{}^{\beta})\eta^{\mu\lambda\nu\sigma}\tilde{R}^{\rho\xi}{}_{\nu\sigma}\nabla_{\xi}\partial_{\mu}\boldsymbol{f}(\phi)  \label{E_GB}, 
\end{eqnarray}
with $\tilde{R}^{\rho\gamma}{}_{\mu\nu} = 
\epsilon^{\rho\gamma\sigma\tau}R_{\sigma\tau\mu\nu}/\sqrt{-g}$, and  
$\dot{\boldsymbol{f}} = \frac{d\boldsymbol{f}}{d\phi}$.
We shall refer to 
$(E_{\alpha}{}^{\beta})\!_{_{_{S\!F\!G\!B}}}$ 
as the Scalar field-Gauss-Bonnet (SFGB) tensor, since it represents
the contribution to the spacetime curvature due to the effects of the SFGB term. 
In fact, $(E_{\alpha}{}^{\beta})\!_{_{_{S\!F\!G\!B}}}$ can be written as, $(E_{\alpha}{}^{\beta})\!_{_{_{S\!F\!G\!B}}} = (E_{\alpha}{}^{\beta})\!_{_{_{S\!F}}} + (E_{\alpha}{}^{\beta})\!_{_{_{G\!B}}}$ where $(E_{\alpha}{}^{\beta})\!_{_{_{S\!F}}}$ is given by (\ref{E_SF}) whereas $(E_{\alpha}{}^{\beta})\!_{_{_{G\!B}}}=- \frac{1}{2}( g_{\alpha\rho} \delta_{\lambda}{}^{\beta} + g_{\alpha\lambda} \delta_{\rho}{}^{\beta})\eta^{\mu\lambda\nu\sigma}\tilde{R}^{\rho\xi}{}_{\nu\sigma}\nabla_{\xi}\partial_{\mu}\boldsymbol{f}(\phi).$  
The structure of the field equations (\ref{modifEqs})
motives the definition of the effective energy-momentum tensor, $\mathscr{E}_{a}{}^{b}$, as $\mathscr{E}_{a}{}^{b} = (E_{a}{}^{b})\!_{_{_{S\!F\!G\!B}}} +  (E_{a}{}^{b})\!_{_{_{N \! L \! E \! D}}}$, thus, the EsGB-NLED theory can be written in a GR-like form, 
$G_{a}{}^{b} = 8\pi\mathscr{E}_{a}{}^{b}$. Therefore,
by taking the divergence of Eq.(\ref{modifEqs}),
and taking into account that 
$\nabla_\beta (E_{\alpha}{}^{\beta})\!_{_{_{N\! L \! E \! D}}}=0$, and 
$\nabla_{\beta} G_{\alpha}{}^{\beta}=0$, from Bianchi identities, it follows that 
$\nabla_\beta(E_{\alpha}{}^{\beta})\!_{_{_{S\!F\!G\!B}}}=0$.  
Our aim here, is to introduce the SSS-AF metric  
(\ref{non_ELLis_WH})
as a pure-magnetic exact solution in EsGB-NLED gravity with a real  scalar field  in the whole T-WH spacetime, i.e. $\phi(r)\in\mathbb{R}$ for all $r\geq|q|$.
%
%
\subsection{EsGB-NLED field equations for the SSS pure-magnetic configuration}
For the line element (\ref{SSSmet}), the non-null 
components of the Einstein tensor are given by (\ref{GabSSS}), 
The non-vanishing 
components of the 
SFGB tensor 
with arbitrary coupling function $\boldsymbol{f}(\phi)$ and potential
$\mathscr{U}$
are
\begin{eqnarray}\label{einstensors}
&& 8\pi(E_{t}{}^{t})\!_{_{_{S\!F\!G\!B}}} = -\frac{ e^{ -2B} }{ 4r^{2} }\left\{  \left[r^{2}e^{B} + 16(e^{B} - 1)\ddot{\boldsymbol{f}} \right]\phi'^{2} - 8[ (e^{B} - 3)B'\phi' - 2(e^{B} - 1)\phi'']\dot{\boldsymbol{f}} \right\} - \mathscr{U}, \label{GBtt} \\
&& 8\pi(E_{r}{}^{r})\!_{_{_{S\!F\!G\! B}}} = \frac{ e^{ -B} \phi'}{ 4 } \left[ \phi'  - \frac{8(e^{B} - 3)e^{-B}A'\dot{\boldsymbol{f}} }{r^{2}}  \right] - \mathscr{U},  \label{GBrr} \\\label{einstensors1}
&& 8\pi(E_{\theta}{}^{\theta})\!_{_{_{S\!F\!G\! B}}} = (E_{\varphi}{}^{\varphi})\!_{_{_{S\!F\!G\! B}}} = -\frac{ e^{ -2B} }{ 4r } \left\{ ( re^{B} - 8A'\ddot{\boldsymbol{f}})\phi'^{2} - 4\left[ (A'^{2} + 2A'')\phi' + (2\phi'' - 3B'\phi')A'\right]\dot{\boldsymbol{f}} \right\} - \mathscr{U}.\label{einstensors2}
\end{eqnarray}
Finally, the  energy-momentum tensor components 
for NLED, 
assuming the SSS spacetime with metric (\ref{SSSmet}), the pure-magnetic field (\ref{magnetica}), and a Lagrangian density $\mathcal{L}(\mathcal{F})$, are given by (\ref{E_nled}).
After replacing the components (\ref{einstensors}-\ref{einstensors2}) in the field equations (\ref{modifEqs}), 
we obtain:
\begin{eqnarray}
&&\!G_{t}{}^{t}\!=\!8\pi\mathscr{E}_{t}{}^{t}\!\hskip.2cm\Rightarrow\hskip.2cm 
4e^{B}\!\!\left( rB' \!+\! e^{ B} \!-\! 1 \right) \!=\!\!\left[ r^{2}e^{B}\!+\!16(e^{ B}\!-\!1)\ddot{\boldsymbol{f}} \right]\!\!\phi'^{2} \!-\!8\!\left[ (e^{ B} \!-\!3)B'\phi' \!-\! 2(e^{ B} \!-\! 1)\phi'' \right]\!\!\dot{\boldsymbol{f}} 
\!+\! 4r^{2}e^{2B}(\mathscr{U}\!+\!2\mathcal{L}), \label{Eqt}\\
&&\nonumber\\
&&\!G_{r}{}^{r}\!=\!8\pi\mathscr{E}_{r}{}^{r}\!\hskip.115cm\Rightarrow\hskip.115cm 4e^{B}\!\!\left( -rA'\!+\! e^{ B} \!-\! 1 \right) \!=\! - r^{2}e^{B}\phi'^{2} \!+\! 8(e^{ B} \!-\! 3)A'\phi'\dot{\boldsymbol{f}} \!+\! 4r^{2}e^{2B}(\mathscr{U}\!+\!2\mathcal{L}),
\label{Eqr}\\
&&\nonumber\\
&&\!G_{\theta}{}^{\theta}\!=\!8\pi\mathscr{E}_{\theta}{}^{\theta}\!\hskip.115cm\Rightarrow\hskip.115cm e^{B}\!\!\left\{ rA'^{2} \!-\! 2B' \!+\! (2 \!-\! rB')A' \!+\! 2rA'' \right\} \!=\! -re^{B}\phi'^{2} \!+\! 8A'\ddot{\boldsymbol{f}}\phi'^{2} \nonumber \\
&& \hskip6.9cm \!+ 4\!\left[ (A'^{2} \!+\! 2A'')\phi' \!+\! (2\phi'' \!-\! 3B'\phi')A' \right]\!\!\dot{\boldsymbol{f}} \!-\! 4 r e^{2B}(\mathscr{U}\!+\!2\mathcal{L}\!-\!4\mathcal{F}\mathcal{L}_{\mathcal{F}}).\label{Eqte}
\end{eqnarray}
Whereas the equation of motion for the scalar field 
$\phi$ can be written as
\begin{equation}\label{phi2}
2r\phi'' + (4 + rA' - rB')\phi' + \frac{4e^{-B}\dot{\boldsymbol{f}}}{r} \left[ (e^{B} - 3)A'B' - (e^{B} - 1)(2A'' + A'^{2})\right] - 4r e^{B}\dot{\mathscr{U}} = 0.
\end{equation}
In the case with $\mathscr{U}(\phi)$=$\Lambda$=constant, and $\mathcal{L}(\mathcal{F})$=$0$, the system of equations (\ref{Eqt}), (\ref{Eqr}), (\ref{Eqte}) and (\ref{phi2}), reduces to that for EsGB gravity with a nonminimally coupled massless scalar field in the presence of a cosmological constant, see for instance \cite{Kanti2018}.
\subsection{Pure-magnetic T-WH supported by NLED in  Einstein-scalar-Gauss-Bonnet gravity} 
Let us present now a specific EsGB-NLED model that leads to the T-WH (\ref{non_ELLis_WH}) be a exact pure magnetic solution of the modified gravity field equations. The following set of characteristic 
functions 
$\boldsymbol{f}=\boldsymbol{f}(\phi(r))$, $\mathscr{U}=\mathscr{U}(\phi(r))$ and  $\mathcal{L}=\mathcal{L}(\mathcal{F}(r))$,  
given respectively by
%
%
\begin{eqnarray}
\boldsymbol{f} &=& - \int^{r}_{|q|} W(\chi)Z(\chi)d\chi, \\
\nonumber\\
\mathscr{U} &=& \frac{q^{4}}{3r^{6}} - \frac{q^{6}}{4r^{8}} - \int^{r}_{|q|}  V(\chi)W(\chi) Z(\chi) d\chi, \\
\nonumber\\  
\mathcal{L} &=& -\frac{q^{2}}{2r^{4}} + \frac{4q^{4}}{3r^{6}} + \frac{q^{6}}{8r^{8}} + \frac{2q^{2}(3q^{2} - 2r^{2})(r^{2}-q^{2}) }{r^{9}}W(r)Z(r)
+\frac{1}{2}\int^{r}_{|q|} V(\chi)W(\chi) Z(\chi) d\chi,
\end{eqnarray}
where we use the auxiliary functions $V(r)=\frac{4 q^{4} \left(5 r^{4} - 7q^{2}r^{2} + q^{4}\right)  }{r^{12}}$, $W(r)=\frac{ q^{2} r \mathrm{e}^{^{-\frac{3q^{2}}{2r^{2}}}} }{\sqrt{r^{2} - q^{2}}}$ and $Z'(r)= \frac{ \mathrm{e}^{^{\frac{3q^{2}}{r^{2}}}} }{q^{2}r^{2}}V(r)$,
defines a EsGB-NLED model for the which the metric (\ref{non_ELLis_WH}) together with the scalar field,
%
\begin{equation}\label{SFreal}
\phi(r) = 2\sqrt{ 1 - \frac{q^{2}}{r^{2}} }  
\end{equation}
is a pure magnetic exact solution of the EsGB-NLED field equations (\ref{Eqt})-(\ref{phi2}). For this solution, the  scalar field (\ref{SFreal}), in contrast to (\ref{SF_non_Ellis}), is a real valued function over the entire T-WH (\ref{non_ELLis_WH}) domain $r\geq|q|$, and therefore satisfies the NEC (\ref{NEC_SF}).

\section{ Behavior of null geodesics and capture cross-section for light }\label{Nullgeods}   
%
%
Now, we  study the behaviour of the null geodesics in the  T-WH geometry (\ref{non_ELLis_WH}), using as a guide the equivalent problem of a particle in a potential well. We  work with the metric written in terms of a new radial coordinate defined by $\rho=\pm\sqrt{r^{2}-q^{2} }$, where the plus (minus) sign is related to the upper
(lower) part of the wormhole.
According to \cite{Rindler}, the geodesic motion of a test particle in this geometry is described by the following Lagrangian density,
\begin{equation}\label{Lagra_Sh_Ell}
2\mathscr{L} = g_{_{\alpha\beta}}\frac{dx^{\alpha}}{d\lambda}\frac{dx^{\beta}}{d\lambda} = - e^{\!^{ -\frac{q^{2}}{\rho^{2} + q^{2}} } }\left(\frac{dt}{d\lambda}\right)^{\!\!\!\!^{2}} + \left(\frac{d\rho}{d\lambda}\right)^{\!\!\!\!^{2}} 
+ \left(\rho^{2}+q^{2}\right)\left[ \left(\frac{d\theta}{d\lambda}\right)^{\!\!\!\!^{2}} + \sin^{2}\!\theta \hspace{0.04cm} \left(\frac{d\varphi}{d\lambda}\right)^{\!\!\!\!^{2}}\right],     
\end{equation}
where $\lambda$ represents an affine parameter of the geodesic.  
The equations of motion for the test particle can be derived from the Euler-Lagrange equations, $\frac{d}{d\lambda}\frac{\partial\mathscr{L}}{\partial p^{\alpha} } - \frac{\partial\mathscr{L}}{\partial x^{\alpha}} = 0$ where $p^{\alpha} = \frac{dx^{\alpha}}{d\lambda}$. Additionally, for  geodesic motion of photons, the Lagrangian has to fulfill the condition $\mathscr{L}(x^{\alpha}, p^{\alpha} ) = 0$.  
The Lagrangian density (\ref{Lagra_Sh_Ell}) does not depend explicitly on the variables $t$ and $\varphi$, then, there exists two conserved quantities associated to them:
$\mathsf{E} = \frac{\partial\mathscr{L}}{\partial\left(\!\frac{dt}{d\lambda}\!\right) }= -e^{\!^{ -\frac{q^{2}}{\rho^{2} + q^{2}} } }\frac{dt}{d\lambda}$, and $\ell= \frac{\partial\mathscr{L}  }{\partial\left(\!\frac{d\varphi}{d\lambda}\!\right) } = (\rho^{2}+q^{2})\sin^{2}\!\theta \hspace{0.04cm}\frac{d\varphi}{d\lambda}$, we can call them the energy and the  angular momentum, respectively.  
To study the motion of test particles in the spacetime geometry (\ref{non_ELLis_WH}) it is convenient to use the fact that  the geodesic motion is always confined to a single plane, because the spherical symmetry. Without loss of generality we will  restrict ourselves to the study of equatorial trajectories
in the  $\theta=\pi/2$ plane.\\
With this choice, the equation of motion for photons 
reduces to,
\begin{equation}
\left(\frac{d\rho}{d\lambda} \right)^{2} + \frac{\ell^{2}}{\rho^{2} + q^{2}} - \Bigg[ e^{\!^{ \frac{q^{2}}{\rho^{2} + q^{2}} } } -  1 \Bigg]\mathsf{E}^{2}   = \mathsf{E}^{2},  
\end{equation}
which can be written as $\left(\frac{d\rho}{d\lambda} \right)^{\!\!\!^{2}} + \mathsf{V}_{\!e\!f\!f}(\rho) = \mathsf{E}^{2}$,
with the effective potential $\mathsf{V}_{\!e\!f\!f}(\rho)$ given by,
\begin{equation}\label{VeffSch_nS_Ellis}
\mathsf{V}_{\!e\!f\!f}(\rho) = \frac{\ell^{2}}{\rho^{2} + q^{2}} + \left( 1 - e^{\!^{ \frac{q^{2}}{\rho^{2} + q^{2}} } } \right)\!\!\mathsf{E}^{2}.    
\end{equation}
The last potential goes to zero as $\rho\to\pm\infty$, and we can verify it has three extreme points $\rho_{_{c}} = \{ 0, \rho_{\!_{\pm}} \}$, such that $\frac{d\mathsf{V}_{\!e\!f\!f}}{d\rho}\Big|_{\rho=\rho_{_{c}}}=0$,  with $\rho_{\!_{\pm}}$ given by, 
\begin{equation}
\rho_{\!_{\pm}} = \pm\sqrt{ \frac{\ln\left( \frac{ e q^{2} \mathsf{E}^{2} }{ \ell^{2} }\right) }{\ln\left( \frac{ \ell^{2} }{ q^{2} \mathsf{E}^{2}}\right)}  }|q| \quad\quad\quad\textup{being}\quad\quad\quad \ln(e) = 1.  
\end{equation}
However, $\rho_{\!_{\pm}}$ 
become real only if,
\begin{equation}\label{rho_pm_real}
\mathsf{E}^{2} = \frac{\ell^{2}}{nq^{2}} \quad\textup{where}\quad\quad n\in(1,e).     
\end{equation}
On the other hand, the images of the extreme points of the effective potential are,
\begin{equation}\label{Veff_evaluado}
\mathsf{V}_{\!e\!f\!f}(0) = \frac{\ell^{2}}{q^{2}}+ ( 1 - e )\mathsf{E}^{2},   
\quad\quad\quad\quad \mathsf{V}_{\!e\!f\!f}(\rho_{\!_{\pm}}) = \frac{\ell^{2}}{q^{2}}\left( \frac{q^{2}\mathsf{E}^{2}}{\ell^{2}}  - \ln\!\left(\frac{eq^{2}\mathsf{E}^{2}}{\ell^{2}}\right) \right). 
\end{equation}
In order to determinate if $\rho_{_{c}} = \{ 0, \rho_{\!_{\pm}} \}$ are minimum or maximum of the effective potential, we must  study the behavior of the signs of $\frac{d^{2}\mathsf{V}_{\!e\!f\!f}}{d^{2}\rho}\Big|_{\rho=\rho_{_{c}}}$, finding,
\begin{eqnarray}
&&\frac{d^{2}\mathsf{V}_{\!e\!f\!f}}{d\rho^{2}}\bigg|_{\rho=0} = \frac{2\ell^{2}}{q^{4}}\left( \frac{e q^{2}\mathsf{E}^{2}}{\ell^{2}} - 1 \right)   \label{ddV_rho0} \\
&&\frac{d^{2}\mathsf{V}_{\!e\!f\!f}}{d\rho^{2}}\bigg|_{\rho=\rho_{\!_{\pm}}} = - \frac{4\ell^{2}}{q^{4}} \ln\!\!\left( \frac{e q^{2}\mathsf{E}^{2}}{\ell^{2}}\right) \ln^{3}\!\!\left( \frac{\ell^{2}}{q^{2}\mathsf{E}^{2}}\right). \label{ddV_rhopm}     
\end{eqnarray}


Depending on the energy of the photon, we have three cases for the values of  $\mathsf{E}^{2}$:
\begin{itemize}  
    
    \item[i)] If $\mathsf{E}^{2} = \frac{\ell^{2}}{nq^{2}}$ where $n\in(0,1]$,   
    according to (\ref{rho_pm_real}) the only real critical point is $\rho_{\!_{c}}=0$, which from (\ref{ddV_rho0}) yields $\frac{d^{2}\mathsf{V}_{\!e\!f\!f}}{d\rho^{2}}\Big|_{\rho=0}>0$, implying  $\mathsf{V}_{\!e\!f\!f}(0)$ is a  minimum value of the potential. This, together with the fact $\mathsf{V}_{\!e\!f\!f}(\rho)\rightarrow0$ as $\rho\rightarrow\pm\infty$ imply the potential is negative everywhere. For the photon energy we are dealing with 
    the relation $\mathsf{E}^{2}>\mathsf{V}_{\!_{eff}}(\rho)$ holds, this means all of them can pass above  this effective potential. See Fig. \ref{V_effA}. 
%
%
\begin{figure}
\centering
\epsfig{file=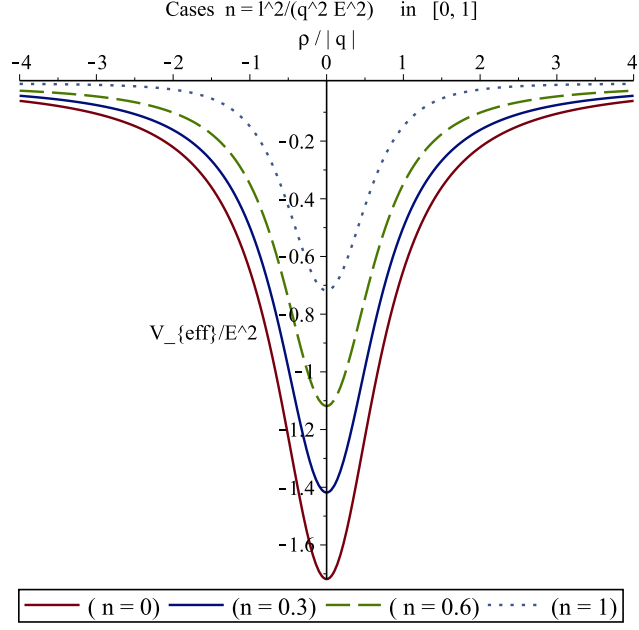, scale=0.44}
\caption{ \label{V_effA} Effective potential for a massless test particle with $\frac{\ell^{2}}{q^{2}\mathsf{E}^{2}}\in(0,1]$, in the spacetime geometry (\ref{non_ELLis_WH}) with $\rho=\pm\sqrt{r^{2} - q^{2}}$. The ordinate is $\frac{\mathsf{V}_{\!e\!f\!f}}{\mathsf{E}^{2}} = \frac{\ell^{2}}{q^{2}\mathsf{E}^{2}}\!\!\left(\frac{\rho^{2}}{q^{2}} + 1\right)^{\!\!\!\!^{-1}} +  1 - e^{\!^{\left(\frac{\rho^{2}}{q^{2}} + 1\right)^{\!\!\!\!^{-1}} }}$; the abscissa is $\rho/|q|$.
}
\end{figure}
\item[ii)] If $\mathsf{E}^{2} = \frac{\ell^{2}}{n q^{2}}$ being $n\in(1,e)$,
    according to (\ref{rho_pm_real}) we have three real critical points in this case, because $\rho_{\!_{\pm}}\in\mathbb{R}$.  From (\ref{ddV_rhopm}) we have $\frac{d^{2}\mathsf{V}_{\!e\!f\!f}}{d\rho^{2}}\bigg|_{\rho=\rho_{\!_{\pm}}}<0$ then $\mathsf{V}_{\!e\!f\!f}(\rho_{\!_{\pm}})$ are local maximum values of the potential, i.e. $\mathsf{V}_{\!e\!f\!f}(\rho_{\!_{\pm}}) = \mathsf{V}^{M\!a\!x}_{\!e\!f\!f}$. On the other hand from  (\ref{ddV_rho0}) we have $\frac{d^{2}\mathsf{V}_{\!e\!f\!f}}{d\rho^{2}}\bigg|_{\rho=0} >0$, implying $\mathsf{V}_{\!e\!f\!f}(0)$ is a local minimum  value of the potential. By looking at the asymptotic behavior   $\mathsf{V}_{\!e\!f\!f}(\rho)\rightarrow0$ as $\rho\rightarrow\pm\infty$, we can conclude $\mathsf{V}^{M\!a\!x}_{\!e\!f\!f}=\mathsf{V}_{\!e\!f\!f}(\rho_{\!_{\pm}})$ is the global maximum value of the potential, whereas $\mathsf{V}_{\!e\!f\!f}(0)$ is the global minimum value. 
    Now, using (\ref{Veff_evaluado}), for $\rho_{
    \!_{\pm}}$ with $\mathsf{E}^{2} = \frac{\ell^{2}}{n q^{2}}$  and $n\in(1,e)$, we have,
\begin{equation}
\mathsf{V}^{M\!a\!x}_{\!e\!f\!f} = \mathsf{V}_{\!e\!f\!f}(\rho_{\!_{\pm}}) =  \frac{\ell^{2}}{nq^{2}} - \frac{\ell^{2}}{q^{2}}\ln\left(\frac{e}{n}\right) \quad\quad \Rightarrow \quad\quad  \mathsf{E}^{2} = \mathsf{V}^{M\!a\!x}_{\!e\!f\!f} + \frac{\ell^{2}}{q^{2}}\ln\left(\frac{e}{n}\right).   
\end{equation}
From the last relation we conclude that for this  photon $\mathsf{E}^{2}>\mathsf{V}^{M\!a\!x}_{\!e\!f\!f}$. because     $\frac{\ell^{2}}{q^{2}}\ln\left(\frac{e}{n}\right)\in\mathbb{R}^{+}$,  and it always can pass above this effective potential. See Fig. \ref{V_effB}.
%
%
\begin{figure}
\centering
\epsfig{file=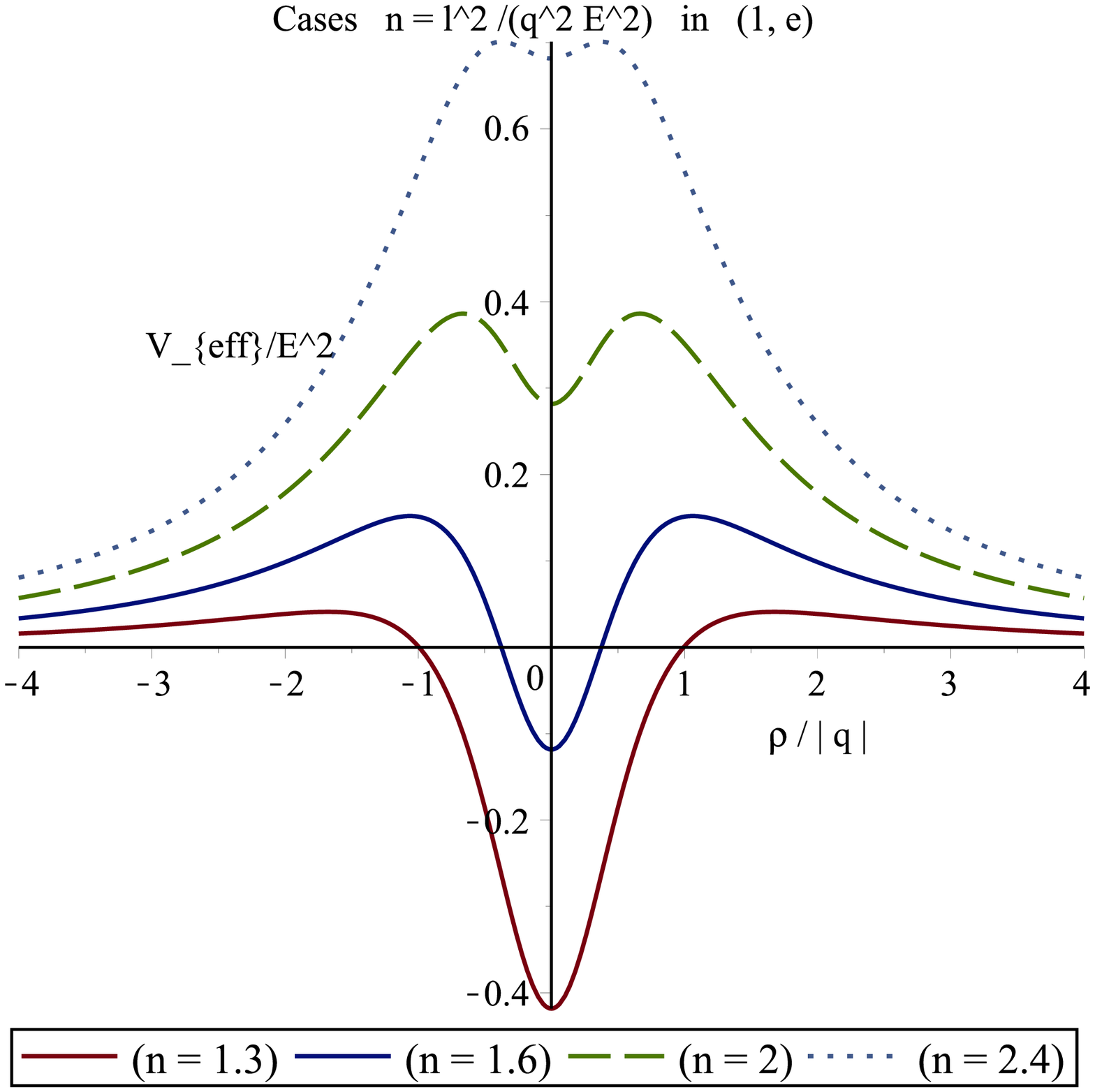, scale=0.44}
\caption{ \label{V_effB} Effective potential for a massless test particle with $\frac{\ell^{2}}{q^{2}\mathsf{E}^{2}}\in(1,e)$, in the spacetime geometry (\ref{non_ELLis_WH}) with $\rho=\pm\sqrt{r^{2} - q^{2}}$. The ordinate is $\frac{\mathsf{V}_{\!e\!f\!f}}{\mathsf{E}^{2}} = \frac{\ell^{2}}{q^{2}\mathsf{E}^{2}}\!\!\left(\frac{\rho^{2}}{q^{2}} + 1\right)^{\!\!\!\!^{-1}} +  1 - e^{\!^{\left(\frac{\rho^{2}}{q^{2}} + 1\right)^{\!\!\!\!^{-1}} }}$; the abscissa is $\rho/|q|$.
}
\end{figure}
%
%
%
\item[iii)] Finally, if 
$\mathsf{E}^{2}=\frac{\ell^{2}}{nq^{2}}$ where $n\in[e,\infty)$, according to (\ref{ddV_rho0}) we have $\frac{d^{2}\mathsf{V}_{\!e\!f\!f}}{d\rho^{2}}\Big|_{\rho=0} = \frac{2\ell^{2}}{q^{4}}\left( \frac{e}{n} - 1 \right) \leq 0 $ this implies  the potential has a local maximum located at $\rho=0$. 
Moreover, since $\mathsf{V}_{\!e\!f\!f}(\rho)\rightarrow0$ as $\rho\rightarrow\pm\infty$ we conclude that $ \mathsf{V}_{\!e\!f\!f}(0) = \mathsf{V}^{M\!a\!x}_{\!e\!f\!f}$ is the maximum value of the effective potential. Now, using (\ref{Veff_evaluado}), for $\rho=0$,
with  $\mathsf{E}^{2}=\frac{\ell^{2}}{nq^{2}}$, we obtain,     
\begin{equation}\label{Veff_c}
\mathsf{V}^{M\!a\!x}_{\!e\!f\!f} = \mathsf{V}_{\!e\!f\!f}(0) = \frac{\ell^{2}}{q^{2}}+ ( 1 - e )\mathsf{E}^{2} = \left( 1 + n - e \right)\mathsf{E}^{2}  \quad\quad \Rightarrow \quad\quad \mathsf{E}^{2} = \frac{\mathsf{V}^{M\!a\!x}_{\!e\!f\!f}}{ 1 + n - e} \leq \mathsf{V}^{M\!a\!x}_{\!e\!f\!f}. 
\end{equation} 
The equation  $\mathsf{E}^{2} = \mathsf{V}^{M\!a\!x}_{\!e\!f\!f}$ can only  be satisfied  for photons with $\mathsf{E}^{2}=\frac{\ell^{2}}{eq^{2}}$, this corresponds to an unstable circular orbit. Whereas, photons with $\frac{\ell^{2}}{q^{2}\mathsf{E}^{2}}\in(e,\infty)$ cannot pass through this effective potential because for them   $\mathsf{E}^{2} < \mathsf{V}^{M\!a\!x}_{\!e\!f\!f}$. 
See Fig. \ref{V_effC}.      
\begin{figure}
\centering
\epsfig{file=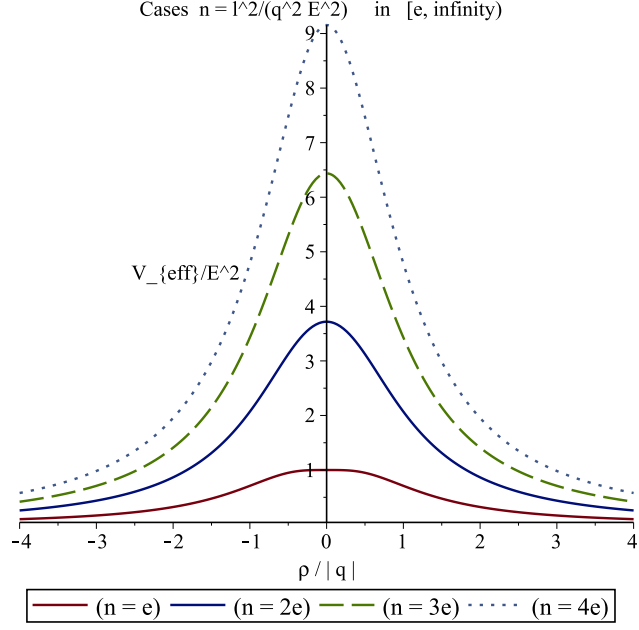, scale=0.44}
\caption{ \label{V_effC} Effective potential for a massless test particle with $\frac{\ell^{2}}{q^{2}\mathsf{E}^{2}}\in[e,\infty)$, in the spacetime geometry (\ref{non_ELLis_WH}) with $\rho=\pm\sqrt{r^{2} - q^{2}}$. The ordinate is $\frac{\mathsf{V}_{\!e\!f\!f}}{\mathsf{E}^{2}} = \frac{\ell^{2}}{q^{2}\mathsf{E}^{2}}\!\!\left(\frac{\rho^{2}}{q^{2}} + 1\right)^{\!\!\!\!^{-1}} +  1 - e^{\!^{\left(\frac{\rho^{2}}{q^{2}} + 1\right)^{\!\!\!\!^{-1}} }}$; the abscissa is $\rho/|q|$.
}
\end{figure}
\begin{center}
    \begin{figure}[htb!]
        \centering
        \includegraphics{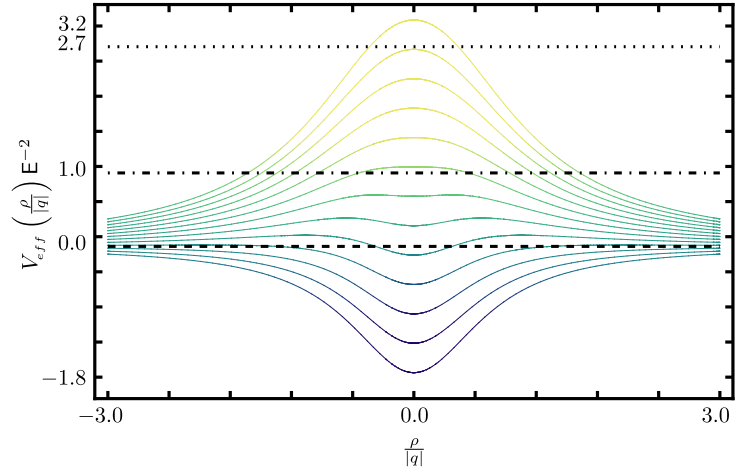}
        \caption{Effective potential curves for several values of $n=\frac{\ell^{2}}{q^{2}\mathsf{E}^{2}}$, horizontal  lines represent  important values of the parameter $n$.}
        \label{fig:my_label}
    \end{figure}
\end{center}
\end{itemize}
{\bf Photon sphere:} We have shown that in the spacetime geometry (\ref{non_ELLis_WH}) is possible that some 
photons to follow circular orbits. Specifically, according to (\ref{Veff_c}), a photon with $\mathsf{E}^{2}=\frac{\ell^{2}}{eq^{2}}$ feel a potential so that  
$\mathsf{E}^{2}=\mathsf{V}_{\!e\!f\!f}(0)=\mathsf{V}^{M\!a\!x}_{\!e\!f\!f}=\frac{\ell^{2}}{eq^{2}}$, 
which implies that  
this could follow a circular geodesic. 
This orbit is called the
photon circle (for details see\cite{pho_orb}). 
Due to the spherical symmetry, the condition $\mathsf{E}^{2}=\mathsf{V}^{M\!a\!x}_{\!e\!f\!f}$ defines a collection of infinitely many such orbits, therefore the last photon orbit
is also called photon sphere \cite{pho_orb}.
According to \cite{Wald}, the impact parameter $\boldsymbol{b}$  will be $\boldsymbol{b} =  \ell/\sqrt{\mathsf{V}_{\!e\!f\!f}^{Max}} = \ell/|\mathsf{E}| = e^{^{1/2}} |q|$, and hence the
capture cross-section for a light beam is $\sigma=\pi\boldsymbol{b}^{2}= e \pi q^{2}$, or in terms of the WH throat, $r=r_{_{0}}=|q|$, this becomes $\sigma= e \pi r^{2}_{_{0}}$.
The photon sphere can cast a wormhole shadow for an observer at infinity.
This shadow is a disk specified by its radius $r_{sh}$ and it gives an apparent size and shape of WH throat. For a static spherically symmetric asymptotically flat wormhole $r_{sh}$ is just the impact
parameter $\boldsymbol{b}$. So for example, for the Ellis WH $r_{sh} = |q|$ which in terms of the Ellis WH throat $r_{_{0}}=|q|$, becomes $r_{sh} = r_{_{0}}$, see \cite{EllisShadows} for details. 
In our case for the WH geometry (\ref{non_ELLis_WH}) we get  $r_{sh} = e^{^{1/2}} |q|=e^{^{1/2}} r_{_{0}}$ which is bigger that of the Ellis WH.

{\bf Asymptotically behavior of the metric (\ref{non_ELLis_WH}) at infinity:}
Expanding the metric in powers of $q/r$ around $r\to\infty$ give us
\begin{equation}
-g_{_{tt}} = e^{\!^{ -\frac{q^{2}}{r^{2}} } } = 1 - \frac{q^{2}}{r^{2}} + \mathcal{O}\!\!\left(\frac{q^{4}}{r^{4}}\right).   
\end{equation}
This allow us to write the line element as,
\begin{equation}
\boldsymbol{ds^{2}} = - \left(1 - \frac{q^{2}}{r^{2}} + \mathcal{O}\!\!\left(\frac{q^{4}}{r^{4}}\right) \right)\boldsymbol{dt^{2}} + \frac{\boldsymbol{dr^{2}}}{1 - \frac{q^{2}}{r^{2}} } + r^{2}\boldsymbol{d\Omega^{2}}, 
\end{equation}
which behaves asymptotically as the exterior region of the  Reissner-Nordstr\"om black hole: $-g_{_{tt}}=1/g_{_{rr}}=1 - 2M_{\!_{A\!D\!M}}/r + Q^{2}/r^{2}$ , without  Arnowitt-Deser-Misner (ADM) mass ($M_{\!_{A\!D\!M}}=0$) and with imaginary charge ($Q = iq$). The latter is a immediate consequence of the violation of the null and weak energy conditions in the region of weak field $\mathcal{F}\rightarrow 0 \equiv r/|q| \rightarrow\infty$ by the NLED Lagrangian density (\ref{NLmax1}). 

\section{Conclusions}
In this work, in the Einstein's gravity context, we have constructed a new SSS-AF T-WH solution which can be interpreted as a non-trivial redshift modification of the Ellis WH. The sources are; a self-interacting phantom scalar field, which was introduced to satisfy the  flare-out condition; and a nonlinear electrodynamics field which becomes the Maxwell theory in the limit of weak field. Moreover, this nonlinear electrodynamics satisfying the WEC in a limited region of the spacetime which contains the WH throat.  We also study the source of the T-WH from the perspective of modified gravity, using a Gauss-Bonnet correction term, wich provide the necessary conditions for the existence of such solution without claiming the existence of a panthom field. Our T-WH metric is determined by only one parameter $q$, which can be associated to the magnetic charged, and defines the WH throat as $r_{_{0}}=|q|$. Thus, in the limit of zero magnetic charge, the Minkowski metric is recovered. Moreover, we found that this solution has a WH shadow of radius $r_{_{sh}} = e^{^{1/2}} r_{_{0}}$ which is bigger than the shadow radius of the Ellis WH. 

To a better characterization of the wormhole we have just presented is necessary the study of their quasinormal modes, its corresponding Penrose diagram and its stability. This last behavior is essential in order to guarantee the traversability of this wormhole.  We hope to return to these issues in a future work.

\appendix 
\section{\bf Null and weak energy conditions in GR}

For a energy-momentum tensor $T_{\mu\nu}$, the null energy condition (NEC),  stipulates that for every null vector, $n^{\alpha}$, yields $T_{\mu\nu}n^{\mu}n^{\nu}\geq0$.
Following \cite{WEC}, for a diagonal energy-momentum tensor $(T_{\alpha\beta})=\mathrm{diag} \left( T_{tt},T_{rr},T_{\theta\theta},T_{\varphi\varphi} \right)$, which can be
conveniently written as, 
\begin{equation}\label{diagonalEab}
T_{\alpha}{}^{\beta} = - \rho_{t} \hskip.05cm \delta_{\alpha}{}^{t}\delta_{t}{}^{\beta} + P_{r} \hskip.05cm \delta_{\alpha}{}^{r}\delta_{r}{}^{\beta} + P_{\theta} \hskip.05cm \delta_{\alpha}{}^{\theta}\delta_{\theta}{}^{\beta} + P_{\varphi} \hskip.05cm \delta_{\alpha}{}^{\varphi}\delta_{\varphi}{}^{\beta} ,
\end{equation}
 where $\rho_{t}$ may be interpreted as the rest energy density of the matter,  
 whereas $P_{r}$, $P_{\theta}$ and $P_{\varphi}$ are respectively the pressures along the $r$, $\theta$ and $\varphi$ directions. In terms of (\ref{diagonalEab}) the NEC implies:
\begin{equation}\label{NEC_Eab}
\rho_{t} + P_{a} \geq0 \quad\quad\textup{with}\quad\quad  a = \{ r, \theta, \varphi \}.
\end{equation}
The weak energy condition (WEC) states that for any timelike vector $\boldsymbol{k} = k^{\mu}\partial_{\mu}$, (i.e., $k_{\mu}k^{\mu}<0$), the energy-momentum tensor  obeys the inequality
$T_{\mu\nu}k^{\mu}k^{\nu} \geq 0$, which means that the local energy density $\rho_{\!_{_{loc}}}= T_{\mu\nu}k^{\mu}k^{\nu}$ as measured by any observer with timelike vector 
$\boldsymbol{k}$ is a non-negative quantity.  
For an energy-momentum tensor of the form (\ref{diagonalEab}), the WEC will be satisfied if and only if,
\begin{equation}\label{WEC}
\rho_{t} = - T_{t}{}^{t} \geq0, \quad\quad\quad\quad \rho_{t} + P_{a} \geq0 \quad\textup{with}\quad  a = \{ r, \theta, \varphi \}.
\end{equation}

\begin{itemize}
\item {\bf NEC and WEC for a self-interacting scalar field $(E_{\alpha}{}^{\beta})\!_{_{_{S \! F }}}$}

Identifying (\ref{E_SF}) with (\ref{diagonalEab}), and using (\ref{EttyErr}), yields,
\begin{eqnarray}
&& 8\pi (\rho_{t})\!_{_{_{S \! F }}} = - 8\pi (P_{\theta})\!_{_{_{S \! F }}} = - 8\pi (P_{\varphi})\!_{_{_{S \! F }}} = \frac{1}{ 4 } e^{ -B} \phi'^{2}  + \mathscr{U},  
\\ 
&& 8\pi (P_{r})\!_{_{_{S \! F }}} = \frac{1}{ 4 } e^{ -B} \phi'^{2} - \mathscr{U}      
\end{eqnarray}
since $(\rho_{t})\!_{_{_{S \! F }}} + (P_{a})\!_{_{_{S \! F }}}=0$ for all $a=\theta$, $\varphi$, the tensor $(E_{\alpha}{}^{\beta})\!_{_{_{S \! F }}}$ satisfies the NEC if  
\begin{equation}\label{NEC_SF}
8\pi(\rho_{t})\!_{_{_{S \! F }}} + 8\pi (P_{r})\!_{_{_{S \! F }}}  = \frac{1}{ 2 } e^{ -B} \phi'^{2} \geq0.
\end{equation}
In addition to (\ref{NEC_SF}) if, 
\begin{equation}\label{WEC_SF}
8\pi (\rho_{t})\!_{_{_{S \! F }}} = \frac{1}{ 4 } e^{ -B} \phi'^{2}  + \mathscr{U} \geq0, 
\end{equation}
holds, the  WEC is satisfied. We can see the WEC is more restrictive than the NEC, this is the reason we only use WEC in our work.
\item {\bf NEC and WEC for the nonlinear electrodynamics field $(E_{\alpha}{}^{\beta})\!_{_{_{N \! L \! E \! D}}}$; pure-magnetic case}

By using (\ref{NLED_EM}) and (\ref{E_nled}),
%
\begin{eqnarray}
8\pi (\rho_{t})\!_{_{_{N \! L \! E \! D}}} = - 8\pi (P_{r})\!_{_{_{N \! L \! E \! D}}} =  2\mathcal{L},   
\quad\quad\quad 8\pi (P_{\theta})\!_{_{_{N \! L \! E \! D}}} = 8\pi (P_{\varphi})\!_{_{_{N \! L \! E \! D}}} =  2(2\mathcal{F}\mathcal{L}_{\mathcal{F}} - \mathcal{L}). 
\end{eqnarray}
since $\rho_{t} + P_{r}=0$, the tensor $(E_{\alpha}{}^{\beta})\!_{_{_{N \! L \! E \! D}}}$ satisfies the NEC if,
\begin{equation}\label{NEC_NLED}
8\pi(\rho_{t})\!_{_{_{N \! L \! E \! D}}} + 8\pi (\rho_{\theta})\!_{_{_{N \! L \! E \! D}}}   = 8\pi(\rho_{t})\!_{_{_{N \! L \! E \! D}}} + 8\pi(\rho_{\varphi})\!_{_{_{N \! L \! E \! D}}}  =  4\mathcal{F}\mathcal{L}_{\mathcal{F}}\geq0.
\end{equation}
in addition to (\ref{NEC_NLED}) if,
\begin{equation}\label{WEC_NLED}
8\pi (\rho_{t})\!_{_{_{N \! L \! E \! D}}} =  2\mathcal{L}\geq0, 
\end{equation}
the WEC is satisfied.
\item {\bf NEC and WEC for the effective energy-momentum tensor $(E_{\alpha}{}^{\beta})\!_{_{_{e\!f\!f\!}}}=(E_{\alpha}{}^{\beta})\!_{_{_{S \! F }}}+ (E_{\alpha}{}^{\beta})\!_{_{_{N \! L \! E \! D}}}$}
\begin{eqnarray}
&&8\pi (\rho_{t})\!_{_{_{e\!f\!f\!}}} = \frac{1}{ 4 } e^{ -B} \phi'^{2}  + \mathscr{U}  + 2\mathcal{L} \\
&&8\pi (P_{r})\!_{_{_{e\!f\!f\!}}} = \frac{1}{ 4 } e^{ -B} \phi'^{2} - \mathscr{U} - 2\mathcal{L} \\
&&8\pi (P_{\theta})\!_{_{_{e\!f\!f\!}}} = 8\pi (P_{\varphi})\!_{_{_{e\!f\!f\!}}} = -\frac{1}{ 4 } e^{ -B} \phi'^{2} - \mathscr{U} + 2(2\mathcal{F}\mathcal{L}_{\mathcal{F}} - \mathcal{L})
\end{eqnarray}
So, the tensor $(E_{\alpha}{}^{\beta})\!_{_{_{e\!f\!f\!}}}$ satisfies the NEC if,
\begin{eqnarray}
&& 8\pi (\rho_{t})\!_{_{_{e\!f\!f\!}}} + 8\pi (P_{r})\!_{_{_{e\!f\!f\!}}} = \frac{1}{ 2 } e^{ -B} \phi'^{2} \geq 0, \label{WEC_effect_la}\\
\nonumber\\
&& 8\pi (\rho_{t})\!_{_{_{e\!f\!f\!}}} + 8\pi (P_{\theta})\!_{_{_{e\!f\!f\!}}} = 8\pi (\rho_{t})\!_{_{_{e\!f\!f\!}}} + 8\pi (P_{\varphi})\!_{_{_{e\!f\!f\!}}} = 4\mathcal{F}\mathcal{L}_{\mathcal{F}}\geq0, \label{WEC_effectlb}
\end{eqnarray}
whereas, in addition to (\ref{WEC_effect_la}) and (\ref{WEC_effectlb}), if,
\begin{equation}
8\pi (\rho_{t})\!_{_{_{e\!f\!f\!}}} = \frac{1}{ 4 } e^{ -B} \phi'^{2}  + \mathscr{U}  + 2\mathcal{L} \geq 0 
\end{equation}
the WEC is satisfied.
\end{itemize}
\section{Domain of validity of the WEC and NEC for the NLED model}\label{WEC_NEC_Model}

To find the region where the WEC and the NEC are valid, let's notice the following:
\begin{equation}\label{NLmax1_EC}
q^{2}\mathcal{F} = \frac{ q^{4}}{r^{4}} \in (0,1], \quad\quad  q^{2}\mathcal{L} 
=  \left( - \frac{1}{2}  + \frac{2}{3}|q^{2}\mathcal{F}|^{^{\frac{1}{2}}} - \frac{1}{8}(q^{2}\mathcal{F})\right)q^{2}\mathcal{F},
\end{equation}
then, 
\begin{equation}\label{NLmax2_EC_a} 
q^{2}\mathcal{L}  
\geq 0 
\quad\quad  \textup{only if}  \quad\quad \frac{92 - 32\sqrt{7}}{9} \leq q^{2}\mathcal{F} \leq \frac{92 + 32\sqrt{7}}{9}. %
\end{equation}
Whereas, 
\begin{equation}\label{LF} 
\mathcal{L}_{\mathcal{F}} = - \frac{1}{2}  + |q^{2}\mathcal{F}|^{^{\frac{1}{2}}} - \frac{1}{4}(q^{2}\mathcal{F}). 
\end{equation}
hence, 
\begin{equation}\label{NLmax2_EC_A} 
\mathcal{L}_{\mathcal{F}} \geq 0 \quad\quad  \textup{only if}  \quad\quad 6 - \sqrt{32} \leq q^{2}\mathcal{F} \leq 6 + \sqrt{32}  
\end{equation}

Thus, according to (\ref{NEC_NLED}), (\ref{WEC_NLED}) and given that for the pure-magnetic $\mathcal{F}$ is positive defined (\ref{magnetica}), the NLED model holds the WEC (i.e. $\mathcal{L} \geq 0$ and $\mathcal{L}_{\mathcal{F}} \geq 0$) 
only if,
\begin{equation}
\frac{92 - 32\sqrt{7}}{9} \leq q^{2}\mathcal{F} \leq 6 + \sqrt{32}
\end{equation}

However, since in the WH domain  $r\in[|q|,\infty)$, or in terms of the electromagnetic invariant $q^{2}\mathcal{F} = \frac{ q^{4}}{r^{4}} \in (0,1]$, then in the wormhole spacetime (\ref{non_ELLis_WH}) the NLED holds the WEC only if,
\begin{equation}
\frac{92 - 32\sqrt{7}}{9} \leq q^{2}\mathcal{F} \leq 1.    
\end{equation}
%
%
%

\section{\bf Field equations of GR-NLED-SF theory}\label{EB_FieldEqs}

In this appendix we include the more general form of the equations of motion for GR-NLED-SF theory that are satisfied by a SSS metric. 

Since the spacetime is static and spherically symmetric, the more general form of the electromagnetic field tensor is given by,
\begin{equation}\label{Fab_SSS}
F_{\alpha\beta} = \left( \delta^{t}_{\alpha}\delta^{r}_{\beta} - \delta^{r}_{\alpha}\delta^{t}_{\beta}\right) F_{tr} + \left( \delta^{\theta}_{\alpha}\delta^{\phi}_{\beta} - \delta^{\phi}_{\alpha}\delta^{\theta}_{\beta}\right) F_{\theta\phi}.       
\end{equation}
Hence, for an arbitrary NLED Lagrangian density $\mathcal{L}(\mathcal{F})$, the non-vanishing components of the NLED energy-momentum tensor, assuming the SSS metric (\ref{SSSmet}) and the more general SSS electromagnetic field tensor (\ref{Fab_SSS}), are given by,
\begin{eqnarray}
&&8\pi (E_{t}{}^{t})\!_{_{_{N \! L \! E \! D}}} = 8\pi (E_{r}{}^{r})\!_{_{_{N \! L \! E \! D}}} = 2( F_{tr}F^{tr}\mathcal{L}_{\mathcal{F}} - \mathcal{L}), \label{NLEDtt}\\
&&8\pi (E_{\theta}{}^{\theta})\!_{_{_{N \! L \! E \! D}}} = 8\pi (E_{\phi}{}^{\phi})\!_{_{_{N \! L \! E \! D}}} = 2(F_{\theta\phi}F^{\theta\phi}\mathcal{L}_{\mathcal{F}} - \mathcal{L}). 
\end{eqnarray}
Inserting the above components in the Einstein field equations
$C_{\alpha}{}^{\beta} = G_{\alpha}{}^{\beta} - 8\pi [ (E_{\alpha}{}^{\beta})\!_{_{_{S \! F }}} + (E_{\alpha}{}^{\beta})\!_{_{_{N \! L \! E \! D}}} ] = 0$, 
for the metric ansatz (\ref{SSSmet}) and scalar field energy-momentum tensor (\ref{EttyErr}), yield, 
\begin{eqnarray}
&&\!C_{t}{}^{t}\!=\!0\hspace{1.4cm}\!\Rightarrow\!\hspace{0.4cm}
\frac{ e^{^{\!\!-B}} }{ r^{2} }\!\!\left( -rB' - e^{^{\!B}} + 1 \right) + \frac{1}{ 4 } e^{ -B} \phi'^{2}  + \mathscr{U} 
- 2( F_{tr}F^{tr}\mathcal{L}_{\mathcal{F}} - \mathcal{L}) =0, 
\label{AB_Eqt_EM}\\
&&\nonumber\\
&&\!C_{r}{}^{r}\!=\!0\hspace{1.4cm}\!\Rightarrow\!\hspace{0.4cm} \frac{ e^{^{\!\! -B}} }{ r^{2} }\!\!\left( rA' - e^{^{\!B}} + 1 \right) - \frac{1}{ 4 } e^{ -B} \phi'^{2} + \mathscr{U}  
- 2( F_{tr}F^{tr}\mathcal{L}_{\mathcal{F}} - \mathcal{L}) = 0,
\label{AB_Eqr_EM}\\
&&\nonumber\\
&&\!C_{\theta}{}^{\theta}\!=\!C_{\varphi}{}^{\varphi}\!=\!0\hspace{0.4cm}\!\Rightarrow\!\hspace{0.4cm} 
\frac{ e^{^{\!\!-B}} }{ 4r }\!\!\left( rA'^{2} - rA'B' + 2rA'' + 2A' - 2B' \right) + \frac{1}{ 4 } e^{ -B} \phi'^{2}  + \mathscr{U}    
- 2(F_{\theta\phi}F^{\theta\phi}\mathcal{L}_{\mathcal{F}} - \mathcal{L}) =0.
\label{AB_Eqte_EM}
\end{eqnarray}

\end{document}